\documentclass[a4paper]{jpconf}
\usepackage{amsmath,amsfonts,amssymb,amsthm,amsbsy,euscript}

\newtheorem{theor}{Theorem}

\theoremstyle{definition}

\newtheorem{lemma}[theor]{Lemma}
\newtheorem{cor}[theor]{Corollary}
\newtheorem{conjecture}[theor]{Conjecture}

\newtheorem{OpenProblem}{
Problem}
\newtheorem{example}{Example}

\theoremstyle{remark}
\newtheorem{rem}{Remark}

\def\oldvec{\mathaccent "017E\relax }
\newcommand{\OrM}{{\rm O\oldvec{r}}}

\newcommand{\BBR}{\mathbb{R}}

\newcommand{\BBS}{\mathbb{S}}
\newcommand{\BBZ}{\mathbb{Z}}

\newcommand{\cP}{\mathcal{P}}\newcommand{\cQ}{\mathcal{Q}}

\newcommand{\cX}{{\EuScript X}}    

\newcommand{\bx}{{\boldsymbol{x}}}

\newcommand{\veps}{\varepsilon}

\newcommand{\dd}{\partial}
\newcommand{\Id}{{\mathrm d}}

\newcommand{\aff}{\text{\textup{aff}}}

\newcommand{\schouten}[1]{[\![#1]\!]}

\DeclareMathOperator{\sym}{sym}

\newcommand{\by}[1]{\textrm{{#1}}}
\newcommand{\jour}[1]{\textit{{#1}}}
\newcommand{\vol}[1]{\textbf{{#1}}}
\newcommand{\book}[1]{\textit{{#1}}}

\begin{document}
\title{Open problems in the Kontsevich graph construction of Poisson bracket symmetries}

\author{Arthemy V Kiselev
}

\address{
Bernoulli Institute for Mathematics, 
Computer Science \& 
Artificial Intelligence, University of Groningen, P.O.~Box~407, 9700~AK Groningen, The~Netherlands}

\ead{a.v.kiselev@rug.nl}

\begin{abstract}
\noindent%
Poisson brackets admit infinitesimal symmetries which are encoded 
using ori\-en\-ted graphs; this construction is due to Kontsevich (1996). We formulate several open problems about combinatorial and topological properties of the graphs involved, about integrability and analytic properties of such symmetry flows (in particular, for known classes of Poisson brackets), and about cohomological, differential geometric, and quantum aspects of the theory.
\end{abstract}

\noindent
\textbf{1. Introduction.}\quad
Whenever a (non)linear ODE system in a finite\/-\/dimensional model from celestial or rigid body mechanics, optimisation and control, optics and photonics, (bio)chemistry, or molecular biology is 
Poisson, $\tfrac{\Id}{\Id t}\bx=\{\bx,H\}_{\cP}$,
the problem of deforming this model into a continuous family (along a new parameter $\veps\in\mathcal{I}\subseteq\BBR$) can be seen in two different ways. By modifying the Hamiltonian function~$H$, one changes the available energy which is transformed into model's evolution by the Poisson bracket. On the other hand, by keeping the Hamiltonian unchanged but by shifting the Poisson bracket~$\{\cdot,\cdot\}_{\cP}$ into a family~$\{\cdot,\cdot\}_{\cP(\veps)}$, one changes the algorithm how the given energy store is processed into dynamics and integrated over the time~$t$ into trajectories. A particular case of the latter is the infinitesimal shift $\cP\mapsto\cP+\veps\lshad\cP,\cX\rshad+\bar{o}(\veps)$ of the Poisson bi\/-\/vector~$\cP$ (hence the bracket) by taking its own Schouten bracket with a globally defined (at all points $\bx\in M^r$ of the model configuration space) vector field~$\cX$. If so, the `novelty' of the shifted Poisson structure is fictitious because this shift of~$\cP$ in fact amounts to an infinitesimal diffeomorphism of~$M^r$ along the integral trajectories of~$\cX$. Integrating to finite~$\veps$ from $\veps_0=0$, one obtains a family of descriptions for the same Poisson model by using new variables $\mathbf{x}^(\veps)
$ instead of the old~$\bx$; neither the Hamiltonian~$H$ nor the bracket $\{\cdot,\cdot\}_{\cP}$ would then change in earnest. Apart from all of the above, the scaling $\tfrac{\Id}{\Id\veps}\cP=\cP$, which yields $\cP(\veps)=\exp(\veps)\cdot\cP\bigr|_{\veps_0=0}$, accelerates or slows down the time pace (for $\veps<0$ and $\veps>0$, respectively).

We are interested in finding infinitesimal symmetries $\cQ(\cP)\in\sym\bigl(\schouten{\cP,\cP}
=0\bigr)$ of the Jacobi identity $\tfrac{1}{2}\schouten{\cP,\cP}=0$ for classical Poisson bi\/-\/vectors~$\cP$ such that the symmetries~$\cQ$ are (\textit{i}) nonlinear in~$\cP$ (otherwise it would be the time rescaling), (\textit{ii}) proper, $\cQ(\cP)\not\equiv0$, in the sense that the shift of~$\cP$ to $\cP+\veps\cQ(\cP)+\bar{o}(\veps)$ is not everywhere vanishing on~$M^r$ as soon as $\cP$~is Poisson and $\schouten{\cP,\cP}=0$, and (\textit{iii}) universal w.r.t.\ all real (or complex) finite\/-\/dimensional affine Poisson manifolds $(M^{r<\infty}_{\aff}$, $\cP)$. The Hamiltonian we keep unchanged.

It is standard that infinitesimal symmetries $\cQ$ of Poisson brackets~$\{\cdot,\cdot\}_{\cP}$ are $\dd_\cP$-\/cocycles w.r.t.\ the Poisson differential $\dd_\cP=\lshad\cP,\cdot\rshad$, whence every coboundary $\dd_\cP(\cX)\in\ker\dd_\cP$ is a cocycles for all vectors~$\cX$ but maybe not \textit{vice versa}. The (non)triviality of infinitesimal symmetries~$\cQ$ as the respective Poisson cocycles, i.e.\ the (non)solvability of the coboundary equation $\cQ(\cP)=\lshad\cP,\cX\rshad + \Diamond\bigl(\cP,\lshad\cP,\cP\rshad\bigr)$ for~$\cX$ and~$\Diamond$ is a separate question (which can be addressed after such a symmetry~$\cQ$ is obtained).


A universal construction of nonlinear proper infinitesimal symmetries $\dot{\cP}=\cQ(\cP)$ by using oriented graphs, as well as by using the Lie algebra morphism~$\OrM$ which takes suitable cocycles $\gamma\in\ker[{\bullet}\!{-}\!{\bullet},\cdot]$ in the unoriented graph complex to Poisson cocycles $\cQ(\cP)=\OrM(\gamma)(\cP)\in\ker\lshad\cP,\cdot\rshad$, were designed by Kontsevich in~\cite{Ascona96}. (See~\cite{OrMorphism} and references therein for a detailed analysis why the graph orientation morphism~$\OrM$ actually works.) Examples of this construction for known graph cocycles~$\gamma_{2\ell+1}$ on $n$~vertices and $2n-2$ edges (namely, $n=2\ell+2=4$, $6$, and~$8$) have been given in~\cite{f16,tetra16}, \cite{sqs17}, and~\cite{OrMorphism}, respectively. Practical calculation of graph cocycles is addressed in~\cite{JNMP2017,WillwacherZivkovic2015Table}; the algorithms to verify the Poisson cocycle factorisation through the Jacobi identity are available from~\cite{JPCS2017}. 
In the fundamental work~\cite{WillwacherGRT}, see also~\cite{RossiWillwacher2014}, Willwacher related the unoriented graph complex to generators of the Gro\-then\-dieck\/--\/Teich\-m\"ul\-ler Lie algebra~$\mathfrak{grt}$.
In what follows, we expect some familiarity with the subject, e.g., on the basis of~\cite{Ascona96,Bourbaki2017} and~\cite{f16,OrMorphism,JPCS2017,Bedlewo,NJR2018}
or the lectures~\cite{IUM2019} available on\/-\/line from the Steklov MI~RAS; the notation is standard.

The aim of this paper is to bring together a set of open problems which relate to and interconnect various properties of the universal infinitesimal symmetries for Poisson brackets and of the unoriented graph cocycles that determine a class of such symmetries under 
the orientation morphism. 

\smallskip\noindent%
\textbf{2. Initial datum: affine Poisson manifold $\bigl(M^{r
}_{\text{\textmd{aff}}},\cP\bigr)$.}\quad
The Kontsevich construction~\cite{Ascona96} works for every finite\/-\/dimensional affine 
Poisson manifold~$M^r_{\aff}$. Although the affine transition functions between local coordinate charts amount to the simplest possible linear operators 
and translations (consider the local coordinate `angle' on the circle $\BBS^1=\BBR/2\pi\BBZ$), the manifold~$M^r_{\aff}$ can of course be topologically nontrivial (like~$\BBS^1$). Moreover, the choice of an affine atlas on an even\/-\/dimensional manifold with a nondegenerate Poisson structure can prevent the attainability of Darboux coordinates. (Indeed, that would likely require a nonlinear change of variables.) On odd\/-\/dimensional Poisson manifolds, there cannot be any Darboux coordinates. 
This is important because the Kontsevich symmetries $\cQ(\cP)$ are typically equal to zero at Poisson structures~$\cP$ with constant or very low polynomial degree coefficients (e.g., at the Kirillov\/--\/Kostant linear brackets or Sklyanin's brackets with quadratic coefficients), which is in contrast with the high\/-\/degree Nambu construction in~\S6 below (cf.~\cite{GrabowskiMarmoPerelomov}).

\begin{OpenProblem}
Describe ample classes of finite\/-\/dimensional \emph{affine} 
Poisson manifolds~$\bigl(M^r_{\aff},\cP\bigr)$.
\end{OpenProblem}

\begin{rem}
The globalisation problem of extending the orgraph construction of a given Kontsevich symmetry $\cQ(\cP)\in\sym(\lshad\cP,\cP\rshad=0)$ to a \emph{smooth} atlas requires extra structures and their correlation with the Poisson bracket (e.g., an affine connection~$\nabla$ on~$M^r$ and invariance of the bi\/-\/vector~$\cP$ with respect to it). The formula of extension $\dot{\cP}=\widehat{\cQ}(\nabla)(\cP)$ would typically involve the curvature terms (which vanish while the manifold is affine).
\end{rem}

\begin{OpenProblem}
Find the smooth\/-\/atlas globalisation(s) of the Kontsevich symmetries
$\cQ_{1:\frac{6}{2}}(\cP)=\OrM(\gamma_3)(\cP)$, 
$\cQ_5(\cP)=\OrM(\gamma_5)(\cP)$, and $\cQ_7(\cP)=\OrM(\gamma_7)(\cP)$, 
found in~\cite{f16}, \cite{sqs17}, and~\cite{OrMorphism} respectively.
\end{OpenProblem}

\smallskip\noindent%
\textbf{3. Why are the graph cocycles so special\,?}\\[1pt]
\textbf{3.1. Are the graph cocycles truly graphs\,?}\quad
In the papers~\cite{MK93
} (see also~\cite{Ascona96} and~\cite{OrMorphism,JNMP2017,NJR2018} for a pedagogical review), Kontsevich introduced the graph complex --\,one of the many
\,-- with parity\/-\/even vertices, with a wedge ordering of parity\/-\/odd edges, and the differential $\Id=[{\bullet}\!{-}\!{\bullet},\cdot]$ produced by the graded commutator of graph insertions into vertices. This direction of research was furthered by Willwacher \textit{et al.}~\cite{DolgushevRogersWillwacher,KhoroshkinWillwacherZivkovic,WillwacherGRT}: in particular, in~\cite{WillwacherZivkovic2015Table} a generating function counts the numbers of nonzero (i.e.\ not equal to minus itself) unoriented graphs with respect to their bi\/-\/grading by the vertex\/-\/edge numbers. Nontrivial unoriented graph cocycles can be built using these --\,partly relevant but largely irrelevant\,-- graphs. Yet we wee that the 
cocycles, which are few compared with the total graph numbers, are clearly non\/-\/generic. 

\begin{OpenProblem}
As soon as the vertex number $n$ is large enough, are the combinatorial properties of graphs~$\gamma_a$ in a cocycle $\gamma=\sum_a c_a\cdot\gamma_a$ well defined, i.e.\ stable over the entire equivalence class $[\gamma]=[\gamma+\Id(\lambda)]$ modulo arbitrary coboundaries from graphs~$\lambda$ on $n-1$ vertices\,? The properties to examine can be the distribution of diameters (over the set of graphs~$\{\gamma_a\}$ forming the cocycle~$\gamma=\sum_a c_a\cdot\gamma_a$) and distribution of vertex valencies.\\[1pt]
$\bullet$\quad What is the large\/-\/$n$ and asymptotic ($n\to\infty$) behaviour of the graph cohomology dimension (primarily, on or near the ray of vertex\/-\/edge bi\/-\/gradings ($n$, $2n-2$), which is relevant to the Poisson bracket symmetries); are the dimensions encoded by their own generating function\,?
\end{OpenProblem}

Let us keep in mind that the realisation of a certain differential complex using unoriented graphs --\,visibly non\/-\/generic but boiled down to very low dimensional subspaces\,-- might actually veil the true nature of cocycles if the chosen realisation is `incidental'.

\smallskip
\noindent\textbf{3.2. Wheels first, what next\,?}\quad 
We recall the established Lie algebra 
isomorphism~\cite{WillwacherGRT} between the unoriented graph cohomology group of our interest and the Gro\-then\-dieck\/--\/Teich\-m\"ul\-ler Lie algebra~$\mathfrak{grt}$ introduced by Drinfeld. (On the other hand, the dgLa of these graph cocycles is mapped by the orientation homomorphism $\OrM(\cdot)(\cP)$ 
to the Lie algebra of Poisson bracket symmetries endowed with the Poisson differential~$\lshad\cP,\cdot\rshad$.) Certain integral formulae relating graph cocycles to the $\mathfrak{grt}$ generators, known explicitly in low orders, are offered in~\cite[\S6]{RossiWillwacher2014}.
The composition of a cocycle in a given bi\/-\/grading, including the structure of the wheel cocycles related to the $\mathfrak{grt}$ Lie algebra, is in principle unknown.

\begin{OpenProblem}
Illustrate the work of Willwacher's isomorphism $\mathfrak{grt}\simeq H^0(\operatorname{Gra})$ at small vertex numbers $n=4$, $6$, $8$, $9$, $10$, $\ldots$, where there are nontrivial cocycles in~$\operatorname{Gra}$.
\end{OpenProblem}

\noindent\textbf{3.3. Are there `many more' symmetries of $\{\cdot,\cdot\}_{\cP}$ than connected graph cocycles\,?}\quad
Both the constructions of the vertex\/-\/blow\/-\/up differential $\Id=[{\bullet}\!{-}\!{\bullet},\cdot]$ and graph orientation morphism $
\gamma=\sum_a c_a\cdot\gamma_a\longmapsto \dot{\cP}=\OrM(\gamma)(\cP)$ are so robust that the connectedness assumption for the graphs $\gamma_a$ without tadpoles 
is in fact excessive, although it is 
ofted adopted in the literature.\footnote{The other traditional such (over)assumption is the strong connectedness, i.e.\ the absence of bottlenecks.} 
For instance, whenever $\gamma^1$, $\ldots$, $\gamma^k\in\ker\Id$ are graph cocycles (not necessarily each on $n^i$ vertices and exactly $\#E(\gamma^i)=2n^i-2$ edges) such that $\sum_i n^i=n$ and $\sum_i \#E(\gamma^i)=2n-2$, then the formal sum $\gamma=\bigsqcup_i \gamma^i$ of disconnected graphs, where the disjoint union~$\sqcup$ is linear over each sum $\gamma^i=\sum_a c^i_a\cdot\gamma^i_a$, itself is a graph cocycle, $\gamma\in\ker\Id$, but now on $n$~vertices and precisely $2n-2$ edges, so that the orientation morphism~$\OrM$ yields a new universal symmetry of Poisson brackets. The differential\/-\/polynomial expression encoding that symmetry $\cQ(\cP)$ is the skew\/-\/symmetrised (w.r.t.\ the bi\/-\/vector arguments) product of expressions each stemming from the respective cocycle~$\gamma^i$.

\begin{cor}
Every Kontsevich's symmetry $\cQ(\cP)=\OrM(\gamma)(\cP)$ can be multiplied by an arbitrary power of differential polynomial $\OrM(\omega)(\cP)$, where $\omega\in\ker\Id$ is a graph cocycle on $N$~vertices and exactly~$2N$ edges.
\end{cor}

The graph number Table~2(even$\bullet$) in~\cite{WillwacherZivkovic2015Table} 
(cf.\ \cite[Tab.~2--3]{JNMP2017}) shows that cells along the ray $(\#V,\#E)=(N,2N)$ are populated by nonzero graphs starting at~$N\geqslant 7$. 

\begin{OpenProblem}
Starting which $N'\geqslant 7$, are there nontrivial cocycles $\omega\in\ker\Id$\,? And starting which $N''\geqslant N'$, are these cocycles orientable (i.e.\ meeting no obstructions~\cite[App.\:A]{JPCS2017}) to nonzero Kontsevich's graphs built from wedges ${\leftarrow}\,{\bullet}\,{\rightarrow}$\,? Give explicit examples of new symmetries $\dot{\cP}=\bigl(\OrM(\omega)(\cP)\bigr)^{m\neq0} \cdot \OrM(\gamma)(\cP)$ for known cocycles~$\gamma$ and new such~$\omega$ on the ray~$(N,2N)$.
\end{OpenProblem}

In the above constructions, we meet $m\in\BBZ$ 
cocycle factors in the disjoint product of graphs.

\smallskip\noindent%
\textbf{4. What are the topological identities in the spaces of Leibniz graphs\,?}\\[1pt]
The Poisson cocycle condition, $\lshad\cP,\cQ(\cP)\rshad\doteq0$ to hold by force of $\lshad\cP,\cP\rshad=0$, is satisfied by Kontsevich's symmetries $\cQ(\cP)=\OrM(\gamma)(\cP)$ because the graph orientation morphism~$\OrM$ yields an explicit solution $\Diamond(\cP,\lshad\cP,\cP\rshad)=\sum_i\OrM(\gamma)(\cP,\ldots,\cP_{i-1},\lshad\cP,\cP\rshad_i,\cP_{i+1},\ldots,\cP)$ for the problem $\lshad\cP,\cQ(\cP)\rshad=\Diamond(\cP,\lshad\cP,\cP\rshad)$ of factorisation via the Jacobi identity for Poisson structures~$\cP$ (see~\cite{Ascona96} and~\cite{OrMorphism}). We observed in~\cite{f16,sqs17} that these realisations by the Leibniz graphs, with the Jacobiator inside, are \emph{not unique}: there exist more solutions, $\Diamond=\Diamond_1\neq\Diamond_2$, etc. In other words, there are identities of the form $\Diamond_2-\Diamond_1\equiv0$ in the spaces of Leibniz graphs: when all the Leibniz rules for the derivatives falling on the Jacobiator are worked out and each Jacobiator expands to the standard sum of three terms, the expansions of initially different Leibniz graphs overlap, having common Kontsevich's orgraphs. (The reverse of this was the core idea in our algorithm to generate an Ansatz with Leibniz graphs for the r.-h.s.\ of factorisation problems $\lshad\cP,\cQ(\cP)\rshad=\Diamond(\cP,\lshad\cP,\cP\rshad)$, see~\cite[\S1.2]{JPCS2017}.)

\begin{OpenProblem}
Let us say 
that two Leibniz graphs are adjacent if their expansions have a Kontsevich orgraph in common. Now, describe the topological structure of the identities $\Diamond_2-\Diamond_1\equiv0$ for known non\/-\/unique solutions of the factorisation problems. Otherwise speaking, describe the meta\/-\/graphs with Leibniz graphs as vertices: what is their genus, diameter, typical cycle length, etc.\,?\\[1pt]
$\bullet$\quad Investigate the saturation curve for the number of Leibniz graphs produced by the iterative algorithm (e.g., see \cite[Table~1]{JPCS2017}): at which stage are the new Leibniz graphs typically leaves vs handles\,? What are the last new Leibniz graphs in this sense\,?
\end{OpenProblem}
   
\smallskip\noindent%
\textbf{5. Are the evolution equations $\dot{\cP}=\cQ(\cP)$ integrable\,?}\\[1pt]
The graph orientation morphism~$\OrM$ yields the symmetry $\cQ(\cP)=\OrM(\gamma)(\cP)\in\sym(\lshad\cP,\cP\rshad=0)$ for every graph cocycle $\gamma\in\ker[{\bullet}\!{-}\!{\bullet},\cdot]$ on $n\geqslant4$ vertices and $2n-2$ edges (see~\cite{Ascona96} and~\cite{OrMorphism,Bedlewo}). The right\/-\/hand sides of evolutionary PDE systems $\tfrac{\Id}{\Id\veps}\cP=\cQ(\cP)$ are nonlinear of polynomial degree~$n$ and, generally speaking, proper, i.e.\ $\cQ(\cP)\not\equiv0$ if $\cP$~is Poisson and $\gamma$~is not a coboundary. Not originating from this graph cohomology, the scaling $\dot{\cP}=\cP$, which amounts to $\cP\mapsto\exp(\veps)\,\cP$, is linear in~$\cP$ and proper.

\begin{OpenProblem}
Are there any \emph{other} nonlinear proper infinitesimal symmetries $\cQ(\cP)\in\sym(\lshad\cP,\cP\rshad=0)$ which are universal with respect to all (affine) Poisson manifolds of arbitrary finite dimension but which are not obtained from the graph cocycles by using Kontsevich's orientation morphism~$\OrM$\,?
\end{OpenProblem}

The evolution $\dot{\cP}=\OrM(\gamma)(\cP)$ is formally integrable in terms of (un)oriented graphs: $\cP(\veps)=\exp\bigl(\veps\OrM(\gamma)(\cP)\cdot\vec{\dd}/\dd\cP\bigr)\bigl(\cP\bigr|_{\veps=0}\bigr)$.

\begin{OpenProblem}
Does the formal power series expansion $\cP(\veps)$ using graphs determine a globally well defined Poisson structure at all points of arbitrary affine Poisson manifold $(M^r_{\aff},\cP)$\,? In other words, do the power series converge (uniformly over~$M^r_{\aff}$)\,?
\end{OpenProblem}

The PDE systems $\dot{\cP}=\cQ(\cP)$ which we know from~\cite{tetra16,f16,OrMorphism,sqs17} consist of $r(r-1)/2$ evolution equations with differential\/-\/polynomial right\/-\/hand sides.

\begin{OpenProblem}
What are the (Liouville, in particular super-)\/integrability properties of these nonlinear evolutionary PDE systems $\dot{\cP}=\cQ(\cP)$\,?
\end{OpenProblem}

\begin{conjecture}
Natural mathematical structures themselves usually admit a structure of the same kind.
\end{conjecture}

\begin{OpenProblem}
Does the evolution $\dot{\cP}=\cQ(\cP)$ of Poisson structures~$\cP$ obey the above meta\/-\/principle implying that this evolution itself ought to be Poisson, $\cQ(\cP)=\boldsymbol{\pi}\bigl(\boldsymbol{\chi}(\cP)\bigr)$, where $\boldsymbol{\pi}$~is a variational Poisson structure on the jet space of Poisson bi\/-\/vector components and $\boldsymbol{\chi}$ is a variational covector (possibly, the variational derivative $\delta\mathcal{H}/\delta\cP$ of a well defined Hamiltonian functional, cf.~\cite{TwelveLectures,Olver})\,?
\end{OpenProblem}

\begin{rem}
Another realisation of this meta\/-\/principle could be that the graded Poisson structure $\boldsymbol{\pi}=\lshad\cdot,\cdot\rshad$ is the Schouten bracket and the ``$1$-\/form'' $\boldsymbol{\chi}$ with respect to bi\/-\/vectors~$\cP$ is a $[(-1)+2]$-\/vector field~$\cX$, so that the Poisson dynamics is $\dot{\cP}=\lshad\cP,\cX\rshad$, see~\S7 below.
\end{rem}

\smallskip\noindent%
\textbf{6. Are the ``generic'' Poisson brackets on $\BBR^3$ preserved by the flows\,?}\\[1pt]
\textbf{6.1.}\quad From the work of Nambu and other physicists in the 1970s (cf.~\cite{GrabowskiMarmoPerelomov,ForKac} and references therein) we recall a construction of an ample class of Poisson bi\/-\/vectors $\Id a/\operatorname{dvol}$ on~$\BBR^3$: every such bracket contains two free functional parameters. Denote by $x$, $y$, $z$ the Cartesian (or any other global) coordinates on~$\BBR^3$. Fix a smooth function $a\in C^\infty(\BBR^3)$ and let there be another functional parameter $\varrho(x,y,z)$ referred to the chosen coordinate system. (In earnest, $\varrho$ is a density with nontrivial behaviour under coordinate re\-pa\-ra\-me\-tri\-sa\-ti\-ons, see~\cite{tetra16}.) For any functions $f,g\in C^\infty(\BBR^3)$, put by definition $\{f,g\}_{a,\varrho}(x,y,z) = \varrho(x,y,z)\cdot\det\bigl\|D(a,f,g)\bigr/ D(x,y,z)\bigr\|(x,y,z)$ 
using the Jacobian determinant in the right\/-\/hand side.

\begin{lemma}
The bracket $\{\cdot,\cdot\}_{a,\varrho}$ is Poisson, i.e.\ it is a bi\/-\/linear skew\/-\/symmetric bi\/-\/derivation satisfying the Jacobi identity.
\end{lemma}

\begin{OpenProblem}\label{PrbCovering}
Is the class of Poisson brackets $\{\cdot,\cdot\}_{a,\varrho}$ preserved by any of Kontsevich's infinitesimal symmetries $\dot{\cP}=\cQ(\cP)$\,?
Specifically, for a given symmetry $\dot{P}{}^{ij}=Q^{ij}\bigl[P^{\alpha\beta}([a],\varrho)\bigr]$ of the Poisson bi\/-\/vector $\cP=(P^{ij})$, does there exist a well defined evo\-lu\-ti\-o\-na\-ry system $\dot{a}=A\bigl([a],[\varrho]\bigr)$, $\dot{\varrho}=R\bigl([a],[\varrho]\bigr)$ such that this vector field in the jet space with fibre variables $a$, $\varrho$ is pushed forward by the tangent map~$\tau_*$ to the symmetry $\cQ\bigl[\cP([a],\varrho)\bigr]$, here $\tau\colon(a,\varrho)(x,y,z)\mapsto\{\cdot,\cdot\}_{a,\varrho}\bigr|_{(x,y,z)}$ is the linear differential operator.
\end{OpenProblem}

Note that the (non)existence of the lift $\dot{\cP}\to\dot{a},\dot{\varrho}$ can depend on a choice of the symmetry~$\cQ$.

\smallskip
\noindent\textbf{6.2. Which Poisson brackets are \textit{not} shifted\,?}\quad
All symmetry problems are accompanied by the problem of finding invariant solutions.

\begin{OpenProblem}\label{PrbInvariant}
Describe those two\/-\/parameter Poisson brackets $\{\cdot,\cdot\}_{a,\varrho}$ on~$\BBR^3$ which stay invariant under a given Kontsevich's infinitesimal symmetry: $\cQ\bigl(\{\cdot,\cdot\}_{a,\varrho}\bigr)\equiv 0$ at all $(x,y,z)\in\BBR^3$.\\[1pt]
$\bullet$\quad More generally, describe (classes of) 
bi\/-\/vectors~$\cP$ which are $\cQ$-\/invariant solutions of the Jacobi identity: $\lshad\cP,\cP\rshad=0$ and~$\cQ(\cP)=0$ for a given~$\cQ$.
\end{OpenProblem}

Problems~\ref{PrbCovering}\/--\/\ref{PrbInvariant} are typical in the geometry of differential equations (see~\cite{Olver} and~\cite[Lect.\:3,7]{TwelveLectures}): a symmetry is induced by the projection from a differential covering, and a PDE system is consistently constrained by its own symmetry. 

\smallskip\noindent%
\textbf{7. Are the universal symmetries $\cQ(\cP)\in\ker\lshad\cP,\cdot\rshad$ (non)trivial Poisson cocycles\,?}\\[1pt]
\textbf{7.1.}\quad The $\dd_\cP$-\/triviality of a given Poisson cocycle $\cQ(\cP)$ by definition means the existence of a global vector field~$\cX\in\Gamma(TM^r)$ and, for the construction at the level of Kontsevich's orgraphs, of a sum~$\Diamond$ of Leibniz graphs such that $\cQ(\cP)=\lshad\cP,\cX\rshad+\Diamond(\cP,\lshad\cP,\cP\rshad)$. The term $\Diamond(\cdot,\cdot)$ is invisible if $\cP$~is Poisson. It can be seen (\cite{f16,Ascona96}) that for the Kontsevich tetrahedral cocycle $\gamma_3\in\ker\Id$ and symmetry $\dot{\cP}=\OrM(\gamma_3)(\cP)$, there is no graph~$\delta$ such that $\OrM(\delta)(\cP)$ would be the trivialising vector field~$\cX$. Nevertheless, in all the examples of Poisson bi\/-\/vectors~$\cP$ on affine manifolds~$M^r_{\aff}$, all the symmetries $\cQ(\cP)\in\sym(\lshad\cP,\cP\rshad=0)$ known so far are spon\-ta\-ne\-ous\-ly trivial, i.e.\ they always appear to be $\dd_{\cP}$-\/coboundaries 
w.r.t.\ seemingly \textit{ad hoc} vector fields~$\cX$. (In particular, see~\cite{Anass1702} for illustration and \cite{GrabowskiMarmoPerelomov,LaurentGengouxPichereauVanhaecke} for a store of Poisson brackets.)

\begin{OpenProblem}
Do there exist nonlinear, proper, and still $\dd_\cP$-\/nontrivial infinitesimal symmetries $\cQ(\cP)\in\sym(\lshad\cP,\cP\rshad=0)$ 
of Poisson bi\/-\/vectors~$\cP$ on finite\/-\/dimensional affine Poisson manifolds~$M^r_{\aff}$\,? If the converse it true, what is then the mechanism of $\dd_\cP$-\/triviality of all these Poisson cocycles (in particular, produced by the graph orientation morphism~$\OrM$ from nontrivial graph cocycles~$\gamma\in\ker\Id$)\,?
\end{OpenProblem}

\begin{rem}
Were the infinitesimal symmetries $\dot{\cP}=\cQ(\cP)$ generically \emph{not} $\dd_\cP$-\/exact, then near all or most of all Poisson systems in Nature (provided that their Poisson structures are sufficiently nonlinear) there would be many almost identical systems with the same parameter space and same energy --\,expressed in terms of these common parameters\,-- but with a different finite\/-\/time behaviour. Do we observe this effect empirically\,? (One contrasts the mutations of specimen vs variability of profiles 
for a given specimen, e.g.\ shapes or colourings.)
\end{rem}

\begin{rem}
Whenever the trivialising vector field~$\cX$ exists, it is not uniquely defined. Indeed, for any function $h\in C^\infty(M^r)$, one has $\lshad\cP,\cX\rshad=\lshad\cP,\cX+\lshad\cP,h\rshad\rshad$ because $\dd_\cP^2=0$, so that the $1$-\/vectors $\cX$ themselves mark the $\dd_\cP$-\/cohomology classes.
\end{rem}

\textbf{7.2.}\quad The frequent, generic, or immanent existence of global vector fields $\vec{X}(\cP)$ trivialising the symmetries $\cQ(\cP)=\lshad\vec{X},\cP\rshad$ would mean that the evolution of coefficients of the Poisson bi\/-\/vector~$\cP$ at a point~$\boldsymbol{a}\in M^r_{\aff}$ --\,actually, to the same extent as the evolution $\dot\Omega=\lshad\vec{X},\Omega\rshad$ of \emph{every} multi\/-\/vector at~$\boldsymbol{a}$\,-- is just the reparametrisation of components of the tensor~$\cP$ (resp., $\Omega$) under the infinitesimal shift along the integral trajectories of the vector field $\vec{X}\in\Gamma(TM^r_{\aff})$.
Specifically, the local description of this diffeomorphism at finite~$\veps$ is the change of coordinates:
\[
\mathbf{x}^{\veps\neq0}_{\text{new}}\bigl(\text{point }\boldsymbol{a}\in M^r_{\aff}\bigr) =
\bx^{\veps=0}_{\text{old}}\bigl(\text{point }\boldsymbol{b}\in M^r_{\aff} \bigr| \exp(\veps\vec{X})(\boldsymbol{b})=\boldsymbol{a}\bigr).
\]
This standard lemma in analysis on manifolds is extended --\,from $\dot{Y}=[\vec{X},Y]$ with
the commutator of two vector fields to $\dot{\cP}=\lshad\vec{X},\cP\rshad$ and $\dot{\Omega}=\lshad\vec{X},\Omega\rshad$ with the Schouten bracket of $1$-\/vector~$\vec{X}$ and bi\/-\/vector $\cP$ or any other multi\/-\/vector~$\Omega$\,-- by the recursive definition of the Schouten bracket $\lshad\cdot,\cdot\rshad$ as the graded Leibniz\/-\/rule extension of the commutator~$[\cdot,\cdot]$ (see~\cite{LaurentGengouxPichereauVanhaecke}).

\begin{rem}
Whenever the coefficients of Poisson bi\/-\/vector~$\cP$ are sufficiently nonlinear and if the differential order of the chosen universal symmetry $\cQ$ does not yet trivialise $\cQ(\cP)(\boldsymbol{a})$ at all points $\boldsymbol{a}$ of the manifold~$M^r_{\aff}$ simultaneously, then the $\dd_\cP$-\/trivialising vector field $\vec{X}(\cP)(\boldsymbol{a})$, if exists, would also be 
non\/-\/constant. Hence its trajectories along the charts of~$M^r_{\aff}$ would also be nonlinear. This construction yields a \emph{nonlinear} change of local coordinates
$\mathbf{x}^{\veps\neq0}_{\text{new}}\rightleftarrows \bx^{\veps=0}_{\text{old}}$ on an \emph{affine} manifold\,!
\end{rem}

\begin{OpenProblem}
For a given 
affine Poisson manifold $(M^r_{\aff},\cP)$, how many universal symmetries --\,taken from the countable set $\OrM(\gamma\in\ker\Id)(\cP)$ produced by the graph orientation morphism~$\OrM$ from the $\mathfrak{grt}$-\/related wheel cocycles and their iterated commutators~\cite{KhoroshkinWillwacherZivkovic,WillwacherGRT}\,-- do not vanish identically as global sections from $\Gamma\bigl(\bigwedge^2 TM^r_{\aff}\bigr)$\,?
\end{OpenProblem}

\begin{OpenProblem}
In the above setting, how ample is the set of trivialising vector fields $\vec{X}=\vec{X}(\gamma\in\ker\Id)\in \Gamma(TM^r_{\aff})$ on a given manifold $\bigl(M^r_{\aff},\cP\bigr)$ with a fixed Poisson structure\,?
Is the vector space of these fields~$\vec{X}$ locally complete, to be sufficient for the approximation of arbitrary vector fields on~$M^r_{\aff}$\,?
\end{OpenProblem}

In brief, are the exponents of the trivialising vector fields $\vec{X}(\gamma,\cP)$ enough to imitate a \emph{smooth} structure on an \emph{affine} Poisson manifold\,?

\smallskip\noindent%
\textbf{8. What is the quantisation of Poisson bracket symmetries\,?}\\[1pt]
\textbf{8.1.}\quad The standard procedure of first quantisation in Poisson vs Quantum Mechanics is the replacement of dynamical functions from the algebra $C^\infty(M^r)$ on a Poisson (typically, symplectic) manifold $\bigl(M^r,\{\cdot,\cdot\}_{\cP}\bigr)$ by the associative noncommutative algebra of operators which act on a suitable complex Hilbert space. The correspondence between the classical Poisson bracket and commutator of operators is $\{\cdot,\cdot\}_{\cP}\mapsto\frac{\boldsymbol{i}}{\hbar}[\cdot,\cdot]$; here $\hbar$~is the Planck constant and $\boldsymbol{i}^2=-1$.

\begin{OpenProblem}
For a given infinitesimal symmetry $\cQ\in\sym\{\cdot,\cdot\}_{\cP}$ which acts nontrivially on the Poisson structure $\{\cdot,\cdot\}_{\cP}$ in a given classical model, so that $\cQ\bigl(\{\cdot,\cdot\}_{\cP}\bigr)\not\equiv0$, is there always a well\/-\/defined post\/-\/quantisation response in either or both the Hilbert space of states or the operator algebra of observables\,?
\end{OpenProblem}

\noindent\textbf{8.2.}\quad The Kontsevich construction of deformation quantisation $\times\mapsto\star=\times+\hbar\,\{\cdot,\cdot\}_{\cP}+\bar{o}(\hbar)$ on affine Poisson manifolds $\bigl(M^r_{\aff},\cP\not\equiv0\bigr)$ extends the usual product~$\times$ in the algebra $C^\infty(M^r_{\aff})$ to an associative noncommutative star\/-\/product~$\star$ in the unital algebra $C^\infty(M^r_{\aff})[[\hbar]]$ of formal power series in the deformation parameter~$\hbar$ (see~\cite{MK97}, also~\cite{Kiev18}); the construction is essentially indifferent to the choice of real or complex number field (cf.~\cite{CattaneoFelder2000}). By adding power series tails to the arguments $f$,\ $g$ of~$\star$, i.e.\ by extending $f\mapsto T(f)=f+\hbar T_1(f)+\bar{o}(\hbar)$ and $g\mapsto T(g)=g+\hbar T_1(g)+\bar{o}(\hbar)$ using a sequence $T=\{T_n$, $n\geqslant1\}$ of finite\/-\/order differential operators encoded by the Kontsevich graphs, one acts on the star\/-\/product by the gauge transformation $T\colon\star\mapsto\star'$; it is induced by the formula $f\mathbin{{\star}'}g=T^{-1}\bigl(T(f)\star T(g)\bigr)$.

\begin{OpenProblem}
Describe and illustrate by examples the action of graph complex on the set of star\/-\/products (in particular, in the context of their gauge transformations $\star\rightleftarrows\star'$) whenever the Poisson bi\/-\/vector $\cP$ in the leading deformation term $\hbar\cP$ of $\star=\times+\hbar\,\{\cdot,\cdot\}_{\cP}+\bar{o}(\hbar)$ is shifted by a symmetry $\cQ(\cP)=\OrM(\gamma\in\ker[{\bullet}\!{-}\!{\bullet},\cdot])(\cP)$ to~$\hbar\cP+\veps\cQ(\hbar\cP)+\bar{o}(\veps)$.
\end{OpenProblem}

\begin{example}
The Kontsevich tetrahedral flow (\cite{Ascona96,Bourbaki2017} and~\cite{f16,tetra16}) perturbs the Poisson bracket by the terms starting at $\hbar^{4-1}=\hbar^{3}$, which shows up nontrivially at ~$\hbar^4$ in the expansion~$\star$ mod~$\bar{o}(\hbar^4)$, now available from~\cite{cpp}.
\end{example}

\begin{OpenProblem}
How does the graph complex act --\,by the symmetries $\dot{\cP}=\OrM(\gamma)(\cP)$\,-- on the Ikeda\/--\/Izawa Poisson $\sigma$-\/model~\cite{Ikeda1994} in which the Feynman path integral expansions of 
correlation functions \cite{CattaneoFelder2000} encode the Kontsevich star\/-\/product $\star=\times+\hbar\,\{\cdot,\cdot\}_{\cP}+\bar{o}(\hbar)$\,?\\[1pt]
$\bullet$\quad Does the $\sigma$-\/model in~\cite{Ikeda1994} have symmetries not originating from the graphs $\gamma\in\ker\Id$\,?
\end{OpenProblem}

The presence of hidden (non)\/gauge degrees of freedom in the Kontsevich $\star$-\/products can be put in the further context of Drinfeld's associators.

\begin{OpenProblem}
Likewise, do the universal symmetries $\dot{\cP}=\cQ(\cP)\in\sym\bigl(\lshad\cP,\cP\rshad=0\bigr)$ persist in the course of the Batalin\/--\/Vilkovisky quantisation, $\bigl(\lshad\cP,\cP\rshad=0\bigr)$ $\Longrightarrow$ $\bigl(\boldsymbol{i}\hbar\,\Delta\cP+\tfrac{1}{2}\lshad\cP,\cP\rshad=0\bigr)$, of Poisson bi\/-\/vectors~$\cP$\,? 
\end{OpenProblem}

\begin{OpenProblem}
Is there a well\/-\/defined --\,and universal w.r.t.\ the set $\cQ\in\sym(\lshad\cP,\cP\rshad=0)$ of graph\/-\/encoded symmetries of Poisson structures~$\cP$\,-- generalisation of the Kontsevich construction, originally on affine Poisson manifolds $\bigl(M^{r<\infty}_{\aff},\cP\bigr)$, to the commutative non\/-\/associative calculus of cyclic words with their topological pair\/-\/of\/-\/pants multiplication $\BBS^1\times\BBS^1\to\BBS^1$ in the Kontsevich formal noncommutative (super)\/geometry (see~\cite{KontsevichCyclic} or~\cite{cyclic17} and references therein)\,?
\end{OpenProblem}


\smallskip
{
\noindent\textbf{Acknowledgements.}\quad
The author is grateful to the Organizers of the 26th In\-ter\-na\-ti\-o\-nal Conference on Integrable Systems and Quantum Symmetries (\textsc{ISQS26}, 8--12~July 2019 at \v{C}VUT Pra\-gue, Czech Republic) 
for warm atmosphere during the event. 
This research is supported by JBI RUG project 106552 and the IH\'ES (in part, by the Nokia Fund).
A part of this work 
was done while 
the author 
was visiting at IM~NASU in Kiev, Ukraine.
The author thanks M.~Kontsevich, 
A.\,G.\,Nikitin, and R.~Buring for helpful discussion.

}

\section*{References}
{

}

\end{document}